# Nano-Opto-Electro-Mechanical Systems

L. Midolo[1], A. Schliesser[1], and A. Fiore[2]

[1]*Niels Bohr Institute, University of Copenhagen, Blegdamsvej 17, DK-2100 Copenhagen, Denmark*

[2]*Department of Applied Physics and Institute for Photonic Integration, Eindhoven University of Technology, PO Box 513, 5600 MB Eindhoven, The Netherlands*

**Abstract**

A new class of hybrid systems that couple optical, electrical and mechanical degrees of freedom in nanoscale devices is under development in laboratories worldwide. These nano-opto-electro-mechanical systems (NOEMS) offer unprecedented opportunities to dynamically control the flow of light in nanophotonic structures, at high speed and low power consumption. Drawing on conceptual and technological advances from cavity optomechanics, they also bear the potential for highly efficient, low-noise transducers between microwave and optical signals, both in the classical and quantum domains. This Progress Article discusses the fundamental physical limits of NOEMS, reviews the recent progress in their implementation, and suggests potential avenues for further developments in this field.

# Contents

# Introduction

Controlling light propagation is one of the most important challenges in optics and photonics, and has direct impact on optical communications (e.g. modulation, optical switching, device and network re-configurability), as well as sensing and imaging (e.g. beam steering). From the general laws of electromagnetism, it is clear that such control can be achieved either by a variation of the refractive index in a given medium, or by a displacement of the physical boundaries between media of different indices. The former is employed, e.g., in electro-optic modulators, whereas the latter is used for beam steering by macroscopic or microscopic mirrors. The refractive index tuning range, provided by the application of electric fields, strain, temperature or carrier injection, is limited to $\Delta n = 10^{-3} - 10^{-2}$ in most materials, which often limits the applicability of these approaches. Additionally, the most effective tuning methods (such as temperature tuning and carrier injection) are inevitably associated with significant static power dissipation. In contrast, mechanical displacements can produce large effects (think of a turning mirror) and, in principle, require energy only for switching to a different state. Electrical actuation is readily obtained by exploiting electrostatic or piezoelectric forces. Miniaturization of motorized mirrors and other optical components has led to the development of micro-opto-electro-mechanical systems (MOEMS), which are at the heart of commercial technologies such as digital-light-processing (DLP) beamers and optical switches[1].

The electrical actuation of a moving part within a light-confining structure (e.g. a waveguide or a cavity) can be used to tune the phase or frequency of the corresponding optical field, producing an effective electro-optic interaction (Fig. 1). Importantly, the exploited interactions are fundamentally reciprocal. In particular, light exerts radiation pressure forces, and can induce displacement of a mechanically compliant mirror it is reflected off. Displacements, in turn, can induce voltages and currents in a piezoelectric material or a charged capacitive transducer. The field of cavity optomechanics has intensely studied the intricate dynamics emerging from this coupling throughout the past decade[2]. Whereas the initial focus has rested on one electromagnetic (i.e. optical or microwave) mode and one mechanical degree of freedom only, recent theoretical and experimental work has also brought out the potential of hybrid systems. In particular, the combination of optical, electronic and mechanical functionality enables a range of novel applications, ranging from electric tunability of optomechanical devices[3], to mechanically mediated signal transduction from the microwave to the optical domain. The latter is particularly attractive, given the (albeit as yet theoretical) opportunity for unity efficiency and zero noise temperature.

Taking such systems to the nanoscale—that is, confining electromagnetic and displacement fields to sub-micrometer dimensions—offers opportunities for dramatically enhanced interaction strength, increased bandwidth, lower power consumption, and chip-scale fabrication and integration. These prospects have triggered a mobilization of both nanophotonics and optomechanics communities towards the realization of such nano-opto-electro-mechanical systems (NOEMS, Fig. 1), in spite of the associated technological challenges. In this Progress Article, we review recent progress in this burgeoning field, with a particular emphasis on the underlying fundamentals, the physical limits to miniaturization and speed they imply, and a representative set of particularly promising applications. Given the large body of activity in this field, we choose to restrict the scope of this article to structures that exploit nanoscale light localization in waveguides and cavities, and refer the reader interested in electrically actuated metamaterials and metasurfaces to another recent review[4].

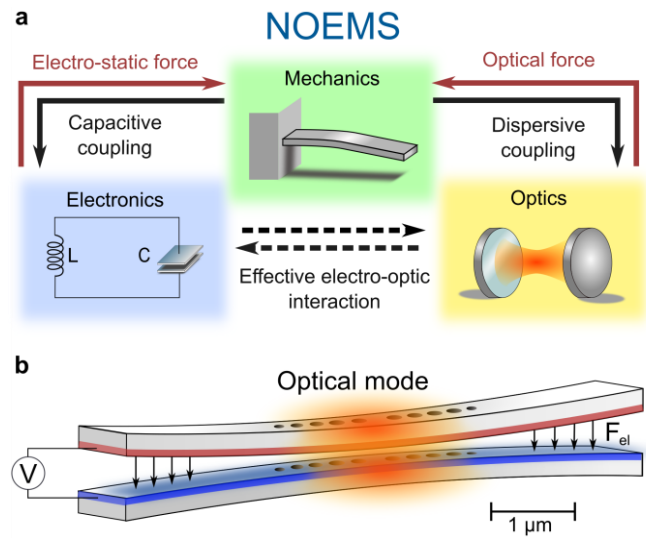

**Figure 1: Physics of nano-opto-electromechanical systems.** (**a**) NOEMS combine three physical systems: electronics, mechanics, and optics. Direct and inverse effects between these systems are mediated by mechanical deformations. In particular, NOEMS allow enhancing electro-optical effects through mechanical degrees of freedom. (**b**) Artistic view of a NOEMS. Electrostatic forces

between two electrodes and optical forces in coupled sub-wavelength waveguides couple charges, mechanical displacement and the optical field.

## Fundamentals of NOEMS

In many photonic materials, the interaction between electrical, mechanical and optical degrees of freedom determines some of the intrinsic properties of solids. For example, the deformation of the atomic lattice under an applied electric field (inverse piezoelectric effect) produces a change in refractive index (photoelastic effect) and thereby contributes to the electro-optic effect. The bulk electro-optic effect depends on the material, but is typically weak in semiconductors and in particular absent in centrosymmetric materials as silicon. NOEMS provide a radically different approach to electro-optic interaction based on a geometrical, rather than intrinsic, effect. They are based on nanomechanical structures designed to respond maximally to an applied electrical force and produce a strong effect on a co-located optical field either through the displacement of their boundaries or through the photoelastic effect (see Box 1). An important example is the case of two parallel and evanescently-coupled nanophotonic waveguides supporting optical modes whose propagation constant depends on the distance between the waveguides and can be actuated electrostatically. Due to the possibility of designing the electro-mechanical and opto-mechanical coupling (Box 1) and the stiffness, such an effective medium can exhibit a strong electro-optic effect regardless of the physical properties of the material of which it is constituted.

A major drive towards reducing opto-electro-mechanical systems to smaller dimensions is given by the fact that opto-electro-mechanical effects become more sizeable at these scales. Optical forces and, in particular, gradient forces, become relevant only in the presence of wavelength-scale confinement and strong gradients of the field, particularly in nano-holes or slots. Similarly, electrostatic forces scale inversely with the square of the charge separation, so that the requirement for high voltage drives is reduced for sub-µm electrode spacing (the actuation voltages for the NOEMS considered here can be reduced to few Volts). Additionally these gaps are shorter than the average distance between electron collisions in air, which allows capacitors to operate without incurring in electrostatic discharges[5], the maximum voltage being ultimately limited by field emission or electromechanical instabilities known as pull-in effect[6].

Another advantage of NOEMS with respect to bulk piezo-electric and photoelastic effects is the possibility to engineer the mechanical response. In the presence of distributed forces, a solid system responds with a deformation which is linear for small deformations (strain within few percent), which holds in most practical situations for crystalline solids. Notwithstanding the complexity of a full three-dimensional displacement function, a generalized Hooke's law of the type $F = kx$ can always be defined for a specific spatial coordinate and a specific load distribution. For simple structures such as cantilevers and doubly-clamped beams, the reduced stiffness $k$ (units of N/m) scales as $\propto EI/L^3$, where $E$ is the Young modulus (a material property), $I$ the moment area of inertia (units of m$^4$) and $L$ the length of the structure. This implies that the stiffness scales linearly when the size of the object is uniformly scaled[4]. When at least one dimension is sub-µm (as in nano-membranes or nano-wires), a spring constant in the order of 1 N/m is easily achievable. Electrostatic forces in capacitive actuators at these scales are in the nN-µN range, allowing deformations up to several tens of nm and correspondingly large optical effects.

NOEMS therefore offer a powerful way to engineer and enhance electro-optic effects in nanophotonic devices. We should, however, mention some notable differences between the electro-optic effect and NOEMS. One important aspect is the response time achievable in these two systems. The electronic response to applied fields is nearly instantaneous so that electro-optic devices are easily operated at 10's of GHz frequencies. This fact is widely exploited for Gb/s data encoding in telecommunication. The electro-mechanical actuation instead, is ultimately limited in speed by the mechanical susceptibility, characterized by a cut-off at the fundamental resonance frequency $\omega = \sqrt{k/m_{eff}}$ where $m_{eff}$ is the effective mass. This represents the equivalent mass that a mechanical mode would have if it were treated as a simple mass-spring system. As for a given force the stiffness is proportional to the displacement, the only solution to achieve faster motion without sacrificing the actuation is to scale the size of the structure (the mass reduces with a third power law, yielding a linear reduction of frequency). Downscaling the structure to sub-µm dimensions allows reducing the switching time to the sub-µs level, well below the ms timescale typical of MEMS. Reaching GHz frequencies requires further scaling the devices to sub-pg masses. This would involve photonic structures with moving parts with dimensions of few tens of nm, and correspondingly a field confinement at these scales, which, for typical near-infrared wavelengths, can only be achieved in plasmonic structures[7,8] or in slotted photonic crystals[9,10]. If repetitive or periodic operation is possible, higher-order modes and resonant driving can be used to reach higher actuation speeds. The mechanical resonances will greatly amplify the motion. Resonant operation is often implemented in Pockels cells to reduce the required driving voltage and to realize pulse-picking and spatial de-multiplexing.

**Box 1 | Optical and electrical forces.**

Electromagnetic forces can be calculated from the Maxwell stress tensor, provided that the electric field $E$ and the magnetic field $B$ are known everywhere in space and that there are no moving charges. However, this requires involved numerical analysis and often is of little practical use. It is much more convenient to treat these forces using the work-energy formalism, where the energy $U$ stored in an electrostatic or optical field gives rise to a force whenever a mechanical motion alters such energy, i.e. $F = -dU/dx$. In non-magnetic materials, only the energy in the electric field is coupled to motion, as the magnetic permeability is constant throughout the structure.

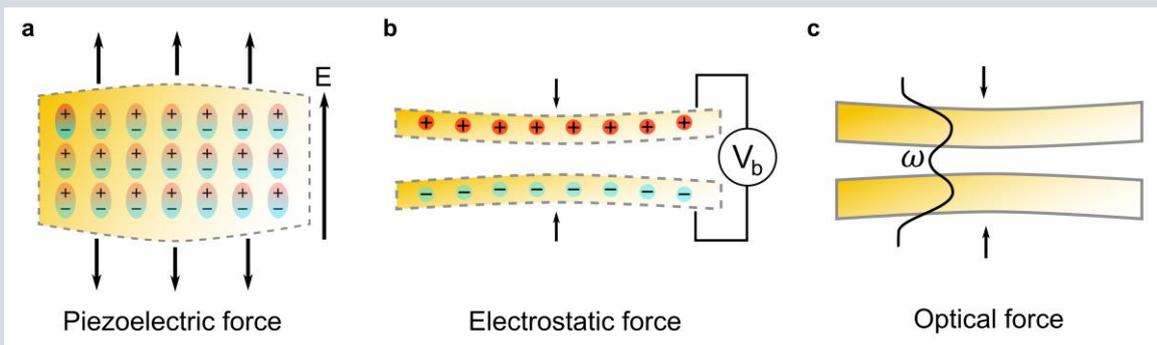

Piezoelectric force     Electrostatic force     Optical force

In a system of fixed charges subject to an external field, as in a piezoelectric material (**a**), the energy can be written as the sum of dipole energies, which depend on distance between charges, corresponding to a force (inverse piezoelectric effect). In the case of an electromechanical capacitor with metal plates (**b**), $U = \frac{1}{2}QV$ (Q and V being the charge on the plates and the voltage between them) the force can be written as: $F = \frac{1}{2} Q \frac{dV}{dx}\big|_Q = \frac{1}{2} V^2 \frac{dC}{dx}$, where *C(x)* is the displacement-dependent capacitance, which is easy to evaluate numerically. For example, in a parallel-plate capacitor of area 10x10 μm², with plates spaced by 200 nm, the force equals ~100 nN under a voltage of 2 V. In the case where energy can be exchanged between the electric and the magnetic field, as in a LC circuit or for optical cavities, the effect of the moving capacitor on the circuit dynamics must be accounted for. However, in the adiabatic limit, the energy exchange, occurring at the electromagnetic resonance frequency *ω*, is much faster than the timescales of mechanical motion produced by the force, so that the electromagnetic system can be seen as a resonator whose frequency is affected by the motion (parametric coupling) (**c**). This guarantees that $U/\omega$ (which corresponds to the number of photons $N_{ph}$) is an invariant (see for example Ref[11]). In this case the force can be written[12] as $F = -N_{ph}\hbar \frac{d\omega}{dx}$ (where $\hbar$ is the reduced Planck constant). This general expression links optical forces to the opto-mechanical coupling factor *dω/dx* which can be calculated from the solution to the Maxwell equations in the optical case[13]. In coupled-nanobeam PhC cavities (Fig. 1b), $\frac{d\omega}{dx} \sim 2\pi\, 100\, GHz/nm$, corresponding to a force of ~66 fN/photon. Note that in both the electrostatic and optomechanical (adiabatic) case the force can be written as $|F| = \frac{U}{L^{eff}}$, where the effective coupling lengths[9] $L_{ES}^{eff} = \left|\frac{1}{C}\frac{dC}{dx}\right|^{-1}$, $L_{OM}^{eff} = \left|\frac{1}{\omega}\frac{d\omega}{dx}\right|^{-1}$ are of order of the dimensions over which the fields are confined (e.g. gap between plates of the capacitor or mirror spacing in a Fabry-Perot cavity) and therefore in the μm- and sub-μm range for NOEMS for both electrostatic and optomechanical actuation. While charges can be confined in sub-μm structures with negligible leakage, it is much more difficult to simultaneously achieve high confinement and small loss rate (thereby high stored energy) for optical fields, so that electrostatic forces tend to be much larger than optical forces for typical operating conditions.

**Box 2 | Enhanced electro-optic effect in parallel waveguides.**

The mechanical deformation of nanophotonic waveguides can be engineered to provide a very strong effective electro-optic interaction in any type of material, including silicon. Here we discuss a specific example, which is at the basis of many NOEMS: a gap-controlled phase shifter. It comprises two closely-spaced parallel waveguides whose distance can be controlled electro-mechanically (**a**). At the core of the phase shifter operation is the splitting of modes into symmetric (S) and anti-symmetric (AS) (or bonding and anti-bonding) supermodes, originating from the evanescent coupling of the individual waveguides. The distance $d$ between the waveguides determines the overlap of the evanescent field of one waveguide with the other one, and therefore the coupling strength μ and the difference in propagation constants between these supermodes according to an exponential law $\propto \exp(-\gamma d)$, where $\gamma$ is the spatial decay of the evanescent field[14]. The gap-dependent splitting translates directly into a variable propagation constant (or effective refractive index) for the two supermodes.

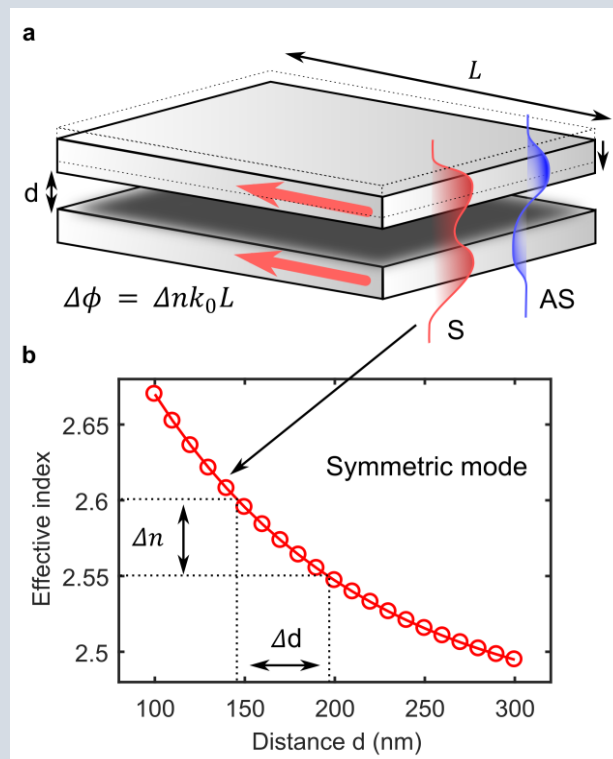

The plot in (**b**) shows the effective index change as a function of the distance for two 160-nm-thick semiconductor slabs (n=3.4) at a wavelength of 1550 nm. The use of electrostatic forces for the motion can lead to a very large electro-optic effect which could be used for phase modulation and switching.

Phase shifters are widely used in photonics, as they form the basis of tuneable lasers and Mach-Zehnder modulators. Phase differences also determine the output of directional couplers, arrayed waveguide gratings and phased arrays. All these systems rely on the controlled variation of optical length resulting in a phase change of $\Delta\phi = \Delta n k_0 L = \pi$, where $k_0 = 2\pi/\lambda$ is the wavenumber in vacuum and $L$ the device length. The relevant figure of merit for a phase shifter is the voltage $V_\pi$ required to obtain a $\pi$-phase shift in a given length. In a NOEMS gap-controlled shifter, a modulation up to $\Delta n_{eff} = 0.05$ and thereby $\pi$ phase shifts in a 15 µm-long waveguide with a distance change of less than 50 nm are possible (**b**). These displacements are typically obtained with less than 10 V in standard capacitors or p-i-n junctions (i.e. the product $V_\pi L \sim 10^{-2}\ V \cdot cm$). The electrostatic nature of the actuation also implies fJ-range actuation energy and nW-level static power dissipation. In crystals such as lithium niobate or PLZT, featuring a relatively high electro-optic coefficient, the small index modulation $10^{-4}$ implies cm-range interaction lengths and $V_\pi L$ products two to three order of magnitude larger than those achievable in NOEMS. In silicon, where the electro-optic effect is absent, static phase modulation is commonly achieved using the thermo-optic effect, which requires large static power dissipation in the range of tens of mW. While a smaller length or index change is needed in resonant devices such as ring resonators[15,16], or in slow-light structures based for example on photonic crystals[17], this comes at the expenses of limited optical bandwidth and high temperature sensitivity, which limits their applicability in real systems.

## Applications to light control and switching

Several applications of NOEMS in nanophotonics have recently emerged, in particular for switching, routing, and phase-shifting in integrated photonic circuits. The main advantages for using mechanics, rather than more conventional electro-optic or thermo-optic effects, are reduced losses, small device footprints, and low-power consumption. Early attempts of using electromechanical actuation for switching relied on controlling the relative alignment between waveguides[18] and sliding reflective structures[19]. These methods however require relatively large displacements (in the range of several µm) and therefore large and complex actuating structures and high applied voltage.

Recently, the attention has shifted to the control of the evanescent coupling between two optical modes (e.g. in two nearby waveguides) by changing their distance[8,20,21]. This relatively simple architecture can be tailored to obtain a plethora of effects, which become stronger in nanophotonic structures due to the large evanescent fields. The simplest, and probably most intuitive, is the change of the propagation constant of the supermodes due to the evanescent coupling (see Box 2).

Experimental demonstrations of MEMS-based switching on silicon have been reported using in-plane motion of directional couplers[22] or ring resonator geometries[23]. Recently, Han et al.[24] and Seok et al.[25] have demonstrated networks of thousands of optical switches based on Silicon directional couplers or adiabatic couplers mounted on electro-mechanical cantilevers (see Fig. 2a) where each switch has very low loss. These examples, although they still involve relatively large micro-mechanical actuators and can therefore be considered as MEMS, demonstrate the great potential of opto-electro-mechanical systems for realizing low-loss networks of switches with MHz-range bandwidth. Moreover they provide interesting solutions and concepts that could be further scaled down in size and optimized for speed. This has been shown by Poot et al[26] using a more compact design of electrodes, where a nano-electro-mechanical phase shifter on SiN waveguides with sub-µs speed has been reported, while a nanomechanical 2x2 switch design with very small actuation voltage and interaction length has been proposed by Liu et al[27]. The next frontier in optical switching will require ~10 ns response times for packet switching. Aggressively scaled nanomechanical systems may manage to achieve these time scales, which would make likely candidates for the switching fabrics in high-performance data center networks.

In the cases discussed above the dispersion relation is still, to good approximation, linear and therefore no group effects are employed. When the dispersion is modified to provide slow-light effects, or optical band-gaps, as in photonic crystals, the mechanical switching can have a dramatic effect on waves with frequencies close to band edges or to a localised resonance. The combination of photonic crystal nano-cavities and nanomechanics has in fact attracted much attention in the recent years. Research in opto-mechanics engineers very strong dispersive couplings $d\omega/dx$ (see Box 1) are engineered in order to enhance radiation pressure, but can also realise higher-order coupling ($\omega \propto x^2$) as required for some sensing protocols[28]. Several works have shown electromechanically-tunable PhC cavities using side-coupled nanowire cavities[29–33], slot waveguides[34], or double-membrane cavities[35]. Some examples are shown in Fig. 2b, 2c, and 2d. Record tuning ranges of up to 30 nm have been obtained with few V applied bias and negligible power dissipation, showing the full potential of electromechanical tuning[36].

Recently, some new applications of mechanical actuation have been explored. Among these, the (electro)mechanical tuning of a photonic structure "on the fly" (i.e. within the photon lifetime) has been proposed as a means to realize frequency conversion[37] and indeed piezoelectric tuning of a waveguide during a single photon's transit can shift the photon's frequency by up to 150 GHz while preserving coherence[38]. Further, rather than controlling the frequency of an optical mode, its optical loss and quality factor can be altered by mechanically modifying the cavity structure[39] or controlling the coupling rate with an output channel such as a waveguide[40–42]. This "dissipative coupling" has been studied in optomechanics as an alternative to the usual dispersive coupling approach, and its electrical control could lead to Q-switched semiconductor lasers and generally to improved control of filters. More generally, a mechanical reconfiguration can be used to modify the field distribution of the cavity mode, leading to modified radiative interactions with integrated quantum emitters[43].

As discussed above, one of the main strength of NOEMS is their compactness and, consequently, the low insertion loss and low power consumption. The benefits of preferring a nano-mechanical approach for optical reconfiguration or switching becomes even more evident in situations where optical amplification is not possible and low-power operation is needed. This is the case, for example, of quantum photonic networks, where the manipulation and routing of single photons (e.g. for boson sampling[44] and quantum simulation[45]) requires reconfigurable architectures, composed of single-photon sources, beam-splitters, phase shifters and detectors. Especially when sources or detectors are integrated on the chip, these circuits require cryogenic operation (< 10 K). As thermo-optic tuning cannot be used at such temperatures and carrier injection produces heating and spurious photon emission, NOEMS are expected to play a key role in quantum photonic networks.

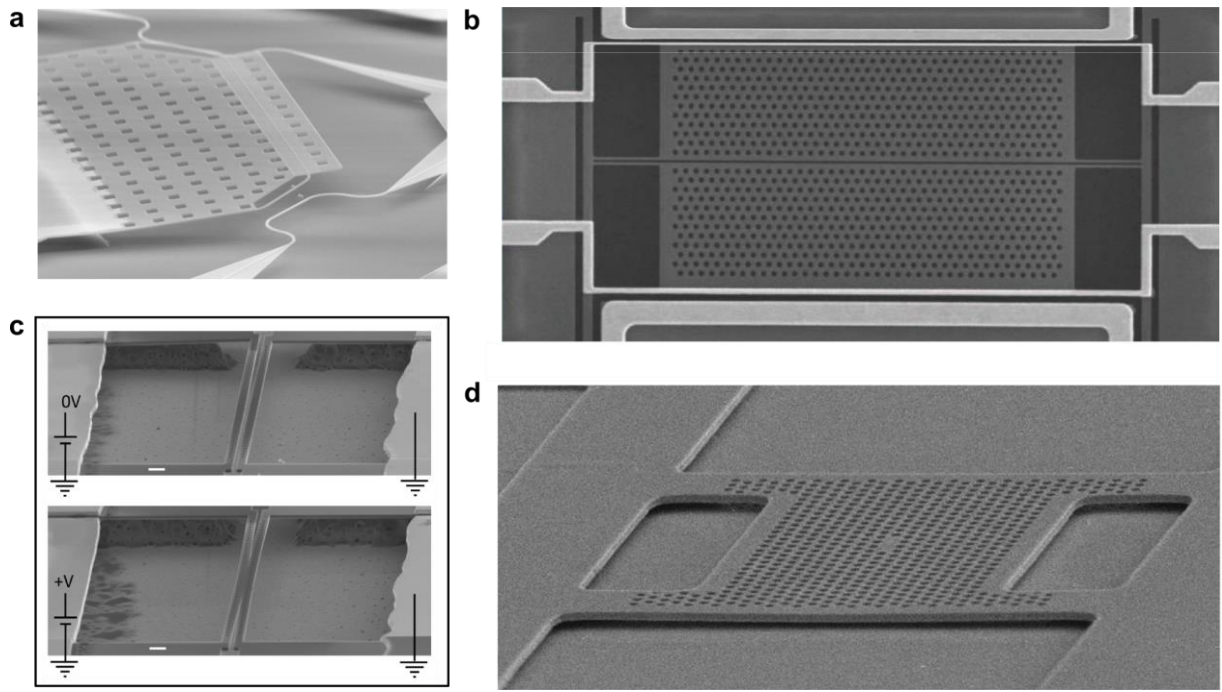

**Figure 2. Examples of NOEMS applications**. (**a**) Controllable optical switch based on micro-electro-mechanical actuation[24]. Light is routed by out-of-plane motion of directional couplers attached to a cantilever. (**b**) An electro-opto-mechanical cavity based on slot waveguides suitable for microwave-to-optical conversion[34,46]. Lateral electrostatic actuators with <50 nm air gaps allow a large wavelength shifts due to the extremely high sensitivity of photonic nanostructures to nano-slots. (**c**) A programmable photonic crystal cavity made of two electrostatically-actuated nanobeams[30]. (**d**) Vertically-actuated 2D photonic crystal cavity on GaAs with embedded quantum emitters[35].

## Applications to signal transduction

Among the most promising applications of the effective electro-optical interaction in NOEMS is the transduction of signals between the electrical and optical domains, using the mechanics as intermediary.

In contrast to coupling via bulk optical nonlinearities, this coupling can be enhanced by tailored mechanical mode shapes, in particular at the nanoscale, as alluded to already above.

Two different regimes of operation can be distinguished, depending on whether a resonance is employed in the electrical and/or mechanical domain (e.g. through the use of an LC circuit). Non-resonant operation can allow the optical detection of electrical signals/charges in a broad frequency range, namely up to the lowest mechanical frequency, which can be in the MHz to 100s MHz range. Considering for example the case of charge detection in a nano-opto-electromechanical PhC cavity, a single electron can produce forces in the fN range, which are easily detected optically with second-range averaging time[47]. The sensitivity can be further boosted by using resident charges to increase the electrostatic force – for example prebiasing the capacitor or using the charges in the depletion region of a p-i-n junction[35]. This may lead to charge sensitivities well below the thermal noise, clearly showing the power of optical sensing of electrical signals and potentially opens the way to optical sensors of electric field or charges, featuring high sensitivity, high spatial resolution and immunity to electromagnetic interference.

Exploiting resonances—both electromagnetic (EM) and mechanical—can dramatically boost the coupling, in particular in conjunction with biasing fields. In the common setting of a parametric coupling, in which mechanical displacements modulate the EM resonance frequencies, the coupling is enhanced: the coupling rate, at which elementary photon-phonon conversion takes place[2], is given by $g = \frac{d\omega}{dx} x_0 \bar{a}$, where $\bar{a}$ is the mean field (normalised such that $|\bar{a}|^2$ is the number of photons in the EM resonator), and $x_0 = \sqrt{\frac{\hbar}{2m\Omega_m}}$ the mechanical mode's zero-point motion. Simultaneously, the biasing fields can fulfil a second crucial role: matching their oscillation frequencies with the differences (or also sums) of mechanical and optical—or electronic—resonance frequencies, renders the parametric coupling effectively resonant, even though the subsystems (optical, mechanical, electronic) reside in very different frequency regimes (100's of THz, MHz, GHz). In a simple picture, a signal conversion process (Fig. 3a) consists of two steps: A microwave cavity photon is converted to the detuned microwave pump frequency through the emission of a phonon. The latter is then upconverted to the optical cavity, assisted by an optical pump photon. This scheme[48–53] extends the coupled photon-phonon dynamics championed by the field of cavity optomechanics[2].

In the ultimate limit, such transducers can be bidirectional and noise-free, enabling high-fidelity conversion of quantum states from the microwave to the optical domain and back[54]. Such a hybrid quantum interface is a crucial, and yet missing, ingredient for networks that connect superconducting qubit processors via optical links[55,56]. The ideal, internal conversion efficiency[54] $\eta = \frac{4C_e C_o}{(C_e+C_o+1)^2}$ of such a transducer is governed by the electro- and optomechanical cooperativies $C_{e/o} = \frac{4g_{e/o}^2}{\kappa_{e/o}\Gamma_m}$, and approaches unity for $1 \ll C_e = C_o \equiv C$. This reflects the competition of couplings $g_{e/o}$ with the loss rates $(\kappa_e, \kappa_o, \Gamma_m)$ of the electric, optical, and mechanical resonators, respectively—but also an impedance-matching condition, favouring matched conversion ($C_e = C_o$). In addition, the mechanics is linked to a thermal bath with a large mean occupation $\bar{n}_{th} \approx k_B T/\hbar\Omega_m$ via its dissipation. The corresponding thermal fluctuations leak into the converter output, resulting in $N \approx \bar{n}_{th}/C \equiv 1/C_q$ noise quanta (per bandwidth per time), where $C_q$ is referred to as the quantum cooperativity. Thus for both key figures of merit, efficiency $\eta$ and added noise $N$, high coupling rates $g_{e/o}$ and small mechanical dissipation $\Gamma_m$ are desirable. A full analysis further must account for external coupling losses (at the input and output of the electromagnetic resonators), parametric amplification of quantum noise, and the resulting performance trade-offs[57].

An early experiment demonstrated measurement of radio-frequency voltage signals via a mechanically resonant membrane transducer[58] whose electrostatically induced out-of-plane motion was detected with a shot-noise-limited laser interferometer (Fig. 3b). Remarkably, it achieved room-temperature voltage sensitivity (<1 nV/√Hz) and noise temperature (<20 K) competitive with state-of-the art electronic amplifiers. Much improved noise performance could be achieved if electronic Johnson noise in the input is reduced; thermomechanical noise and the quantum noise of light (the ultimate limit) add as little as (<60 pV/√Hz) each. Integrated devices of this kind could transduce minute electric signals—for example, from a magnetic resonance scanner—directly to a fibre-carried optical field. The reverse conversion from optical to microwave has also been demonstrated recently[59].

Andrews *et al.*[60] have shown bidirectional, overall 10%-efficient microwave-optical conversion with ~$10^3$ added noise quanta. This system is also based on a SiN membrane, here coupled capacitively to a superconducting LC circuit, and via radiation-pressure with the optical photons in a Fabry-Pérot resonator. Operation at lower temperature, or which more coherent mechanical devices[61], could bring a quantum-enabled transducer into reach. Efforts to downscale such devices are underway in several groups worldwide, promising not only larger coupling rates, but also all-nanofabricated, scalable platforms. For example, working with in-plane mechanical modes of silicon[34,46] or silicon nitride[62] membranes allows the definition and alignment of capacitor electrodes, mechanical structure and optical nanoresonator with nm-scale precision. Sub-100 nm capacitive gaps can be realised in this manner, enabling record coupling rates if parasitic (not mechanically compliant) capacitance is kept at bay.

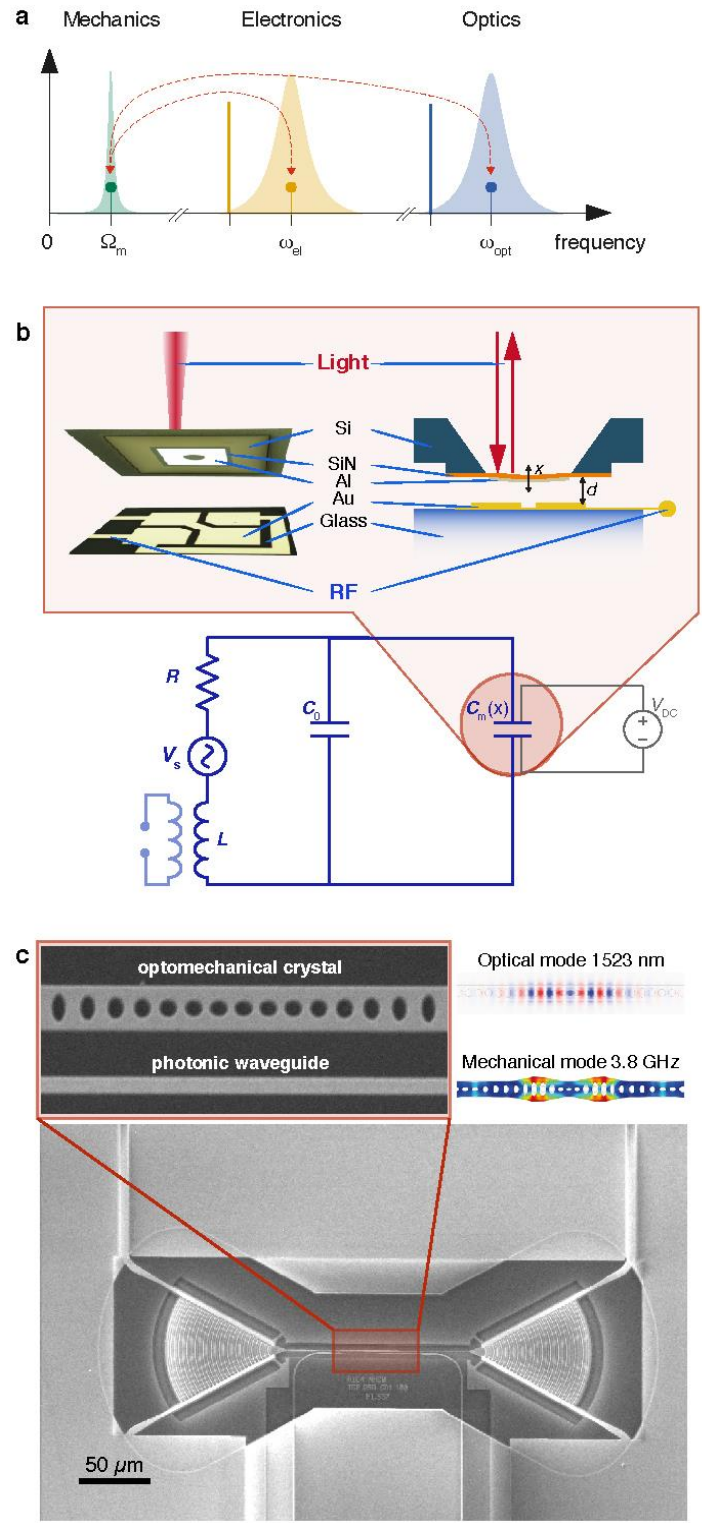

**Figure 3. Opto-electro-mechanical signal transducers**. (**a**) Generic all-resonant signal transducer, coupling excitations (thin lines with disk tip) of a mechanical, electronic, and optical resonance, indicated by green, yellow and blue Lorentzians, respectively. Parametric coupling (dashed red lines) is enhanced by biasing fields (bold yellow and blue lines), tuned to the difference

frequency of the electromagnetic (electronic or optical) and mechanical mode. (**b**) Electro-opto-mechanical transducer for classical radio-frequency signals[58] based on a silicon nitride (SiN) membrane, forming a mechanically compliant capacitor $C_m(x)$. Together with a tuning capacitor $C_0$, and an inductor it forms an RF resonant circuit, in this case degenerate with the mechanical mode, $\omega_{el}=\Omega_m$. Correspondingly, the biasing field is a d.c. voltage. In this proof-of-principle experiment, the optical readout is non-resonant. (**c**) Piezoelectric optomechanical crystal for bidirectional microwave–optical quantum signal conversion[63]. A pair of radially symmetric interdigitated transducers launches Lamb waves towards an optomechanical crystal, which hosts both high quality mechanical and optical modes. The latter can be driven and read out with an optical bias field provided by an evanescently coupled photonic waveguide. Due to the piezoelectric coupling in the device's AlN material, a microwave biasing field is not necessary, and $\omega_{el}=\Omega_m$.

Piezoelectric coupling again provides interesting design alternatives[64] as even high-frequency modes can be efficiently driven without the need to define an electromechanical capacitor. Optomechanical whispering gallery-mode resonators[65] in the piezoelectric AlN[66,67] have been used for early work, followed by several implementations building on optomechanical crystals[9] in the same material. Bidirectional microwave-optical conversion can be achieved by launching GHz surface acoustic waves from an interdigitated transducer[63,68–72] (see Fig. 3c). To date, however, demonstrated "internal" conversion efficiencies are only at the percent level, and lower (order $10^{-4}$) if all in- and output losses are considered[63,68]. Increasing internal efficiency might necessitate a boost in optomechanical coupling, which can hardly come from variations of the highly optimised geometry. It is available in GaAs optomechanical crystals, though, where photoelastic interaction contributes significantly to record-high optomechanical coupling[70,73]. A smaller piezoelectric coefficient is the price to pay in this case, which has, as yet, precluded bidirectional operation with noteworthy efficiency.

From the above examples it is evident that quantum transducers pose extreme demands to the devices' materials and design—even to work in principle, not to mention such practicalities as absorption heating in milliKelvin environments. Yet, it is clear that mechanical transducers are highly promising contenders, given the successes already demonstrated, and known routes for improvement. Direct integration of phononic modes with microwave qubits could improve efficiency in optically addressing the latter[74,75]. Advanced protocols can circumvent stringent requirements on the system's frequency hierarchy[76]. And more options exist for the delicate choice of materials, including large-bandgap piezoelectrics such as GaP. It will be exciting to see the development of these systems, and their performance compared to complementary approaches such as those based on direct electro-optic conversion[77–79] and magnon transducers[80–82], whose conversion efficiencies have as yet remained below the 1%-mark.

Beyond the examples discussed above, a wide range of opportunities has yet to be explored. For example, reservoir engineering or modulation schemes can render signal transport across the microwave and optical spectral domains non-reciprocal[83–85]. This will allow on-chip implementation of isolators and circulators, without the need for magnetic materials. Passive microwave photonic devices, such as filters or delay lines, can be implemented on-chip—with a compact footprint, exploiting the much shorter ($\sim 10^{-5}$) wavelength of phonons compared to electromagnetic waves of the same frequency[73,86]. Optically pumped active devices can provide low-noise microwave amplification, and eventually lead to a new generation of chip-scale microwave oscillators with high spectral purity, as required for advanced communications and radar applications.

## Outlook

The strong effective electro-optic coupling achievable through the nanoscale co-localization of charges, mechanical motion and optical fields makes NOEMS unique contenders for a wide range of applications in communication, sensing and quantum information processing. Progress in theoretical understanding, device design and nanofabrication methods enables the demonstration of increasing functional and efficient structures, ranging from reconfigurable devices and circuits, to fast optical switches, optical sensors and signal transducers. On the route towards turning such concepts into real-world, mass-producible devices, much will hinge on the successful development of suited materials and processes, compatible with CMOS and foundry-level fabrication. Packaging, too, will have to be addressed, given that mechanical system require isolation from the environment including vacuum in some cases. Yet, with a number of major industrial players in the field of microelectronics and MEMS joining this line of research, the prospects are now better than ever.

## Acknowledgements

The authors gratefully acknowledge interesting discussions with N. Calabretta, M. Cotrufo, R.W. van der Heijden, M. Petruzzella, R. Stabile, K. Williams, Z. Zobenica (TU/e), E. Verhagen (AMOLF), P. Lodahl, S. Stobbe (NBI), and K. Srinivasan (NIST). The research leading to these results was funded by the European Union's Horizon 2020 research and innovation programme (ERC project Q-CEOM, grant agreement no. 638765 and FET-proactive project HOT, grant agreement no. 732894), a starting grant and a postdoctoral grant from the Danish Council for Independent Research (grant number 4002-00060 and 4184-00203), and the Dutch Technology Foundation STW, Applied Science Division of NWO, the Technology Program of the Ministry of Economic Affairs under projects Nos. 10380 and 12662.